# The exact description of intruder states in $^{112}$Cd nucleus by using a mixing formalism based on $SU(1,1)$ transitional Hamiltonian and O(6)–Casimir operator

M. Rastegar*, H. Sabri, A. O. Ezzati

Department of Physics, University of Tabriz, 51664 Tabriz, Iran

---

* Corresponding author e-mail: m.rastgar@tabrizu.ac.ir

Abstract;

In this paper, we used a transitional Hamiltonian which has U(5)↔O(6) transition to improve theoretical predictions for energy spectra and quadrupole transition rates of $^{112}$Cd nucleus. To this aim, the transitional Hamiltonian in the affine $SU(1,1)$ algebra has been extended by adding the O(6)-Casimir operator and mixing Hamiltonian to increase exactness in the description of $4_2^+$ and $2_3^+$ intruder levels of this nucleus. We also considered the wave functions of both regular and intruder states, as a combination in the N and N+2 boson spaces. The results confirm the advantages of using such mixing approaches and describing the energy and transition rates with high accuracy.

1. Introduction

One of the genius nuclei that are in the transitional between vibrational and axial symmetry regions are $^{110,112,114}$Cd but in this paper, we want to test just $^{112}$Cd nucleus because it seems great exhibiting for our method and also Lehman in the Ref.[19].emphasize it. Also in Ref [35] mentioned about the particle-hole excitations in $^{108}$Cd by the intruder states that just worked by Hamiltonian. There is a novelty way in this paper that it is work by expansion wave function.

The intruder levels and their observation via experimental techniques, with descriptions of these levels by the theoretical framework, are the subject of different studies. The work of Bohr and Kalkar in nuclear physics has exposed the intention of deformed shapes and the manifestation of disparate shapes in a given nucleus [1]. This idea has been developed by other original studies, such as [2-5], which used different shapes in a nucleus for a complete description of its observables. Various models based on the shell model's concepts, collective geometric model, models based on the mean-field approach with algebraic frameworks are used for such investigations. On the other hand, the multi particles multi holes (mp-mh) is another exclusive method which first proposed in the shell model framework and then expanded in the algebraic models such as the interacting boson model (IBM) to describe different intruder levels in nuclei that are located near the closed proton and neutron shells. The idea of expanding the ordinary 2p-2h intruder excitations into two pairs of extra nucleons (two bosons) and studying their mixing with the regular configurations that first Dual and Barrett presented this approach [2]. It has been widely reported that intruder states have been exposed to low excitation energies [4-10,11,12]. mp-mh excitations cannot be compounded easily in complete large-scale shell-model studies because of the extensive extents of the model spaces involved. The new label has been introduced to represent the symmetries connecting particle and hole bosons within the IBM configuration mixing calculations [13-17]. In this approach, the effect of such excitations is expressed; as different terms of regular Hamiltonian and, therefore; a combination of regular and mixing Hamiltonians e.g.; other symmetries, improve theoretical predictions for both regular and intruder states. A transitional Hamiltonian between different symmetry limits of IBM provides a similar combination of different symmetries, and this idea is the main subject of this investigation.

A detailed investigation of both regular and intruder levels of the Cd isotopic chain has been done by Lehman et al. in Ref.[18]. They have claimed that the vibrational characteristics of $^{112}$Cd portray by the regular configuration subtending of N bosons, s, and d bosons. As a result of the more deformation nature of the intruder configuration, there are two extra bosons involved, N+2, which takes account of a 2p-2h excitation within the closed shells [18-19]. They showed that the $4_2^+$ and $2_3^+$ are intruder states for U(5) algebra, and as anticipated, this limit is too narrow to duplicate the whole set of empirical data, as some low-lying states are outside the model space [20]. To improve the contract between theoretical and experimental results, a subsequent configuration called the intruder configuration had been required. Therefore, they have combined the regular and intruder configurations, which correspond with U(5) and O(6) dynamical limits of IBM, respectively, for N and N+2 bosons spaces.

In this investigation, we extend a similar combination by using a transitional Hamiltonian between U(5) and SO(6) limits which are based on affine $SU(1,1)$ algebra [16-17] as a regular Hamiltonian in the N boson space. In recent studies, this Hamiltonian was used to explain the coexistence of deformed shapes by using the two control parameters of transitional Hamiltonian. Then, this Hamiltonian mixed with the O(6) Casimir operator in the N+2 bosons space to describe intruder levels of $^{112}$Cd nucleus with high accuracy. We also considered the advantages of this mixed formalism in describing of different quadrupole transition rates of this nucleus.

2. Method;

## 2.1. The regular part of Hamiltonian (unperturbed representation):

For the nuclei in the vicinity of the proton closed shell at Z = 50, such as Cd isotopic chain, the regular configuration can be exposed by U(5) algebra in the IBM model, and the intruder configuration gives away the O(6) symmetry [18]. Intruding, the construction owning to the particle-hole excitation amidst the closed shells is a well-known particularity in the nucleus. This configuration is a new implement for the exposition of the level scheme of nuclear that has not been predicted in one algebra. In; cases where the Hamiltonian could be written with the help of the limits of the Casten triangle, the Hamiltonian could be written with invariants operators of two of the three chains [24,25]. One of these transitions is U(5)↔O(6) transition which is used to describe nuclei between vibrational and gamma-unstable limits, nuclei near the closed shells. In the framework of IBM, the even-even Cd isotopes are good candidates to perform the such a transition. This means one may use such transitional Hamiltonian to investigate energy spectra and transition rates of nuclei in this region. The advantages of $SU(1,1)$-based transitional Hamiltonian in comparison with other algebraic approaches which are illustrated in detail in Refs.[17-18]. Encourage us to focus only on this method. We used this $SU(1,1)$ transitional Hamiltonian by the definition of [N] boson presentation; details, about this transitional Hamiltonian, are available in Refs. [17-18,20]. Here, we briefly introduce the main concepts and relations.

The SU(1,1) algebra has been constructed by $S^{\nu}, \nu = 0, \pm$ operators in which the generators $d$ – boson pairing algebra are:

$$S^{+}(d) = \frac{1}{2}(d^{\dagger} \cdot d^{\dagger}) \quad , \quad S^{-}(d) = \frac{1}{2}(\tilde{d} \cdot \tilde{d}) \quad , \quad S^{0}(d) = \frac{1}{4}\sum_{\nu}(d_{\nu}^{\dagger}d_{\nu} + d_{\nu}d_{\nu}^{\dagger}) \tag{1}$$

And these operators for $s$ – boson pairing algebra define as:

$$S^{+}(s) = \frac{1}{2}s^{\dagger 2} \quad , \quad S^{-}(s) = \frac{1}{2}s^{2} \quad , \quad S^{0}(s) = \frac{1}{4}(s^{\dagger}s + ss^{\dagger}) \tag{2}$$

The infinite-dimensional $SU(1,1)$ algebra has been generated by using [17,18]:

$$S_{n}^{\pm} = c_{s}^{2n+1}S^{\pm}(s) + c_{d}^{2n+1}S^{\pm}(d) \quad , \quad S_{n}^{0} = c_{s}^{2n}S^{0}(s) + c_{d}^{2n}S^{0}(d) \tag{3}$$

Where $c_s$ and $c_d$ are real parameters and $n$ can be $0, \pm 1, \pm 2, \ldots$. The commutation relations between these operators are:

$$[S_m^0, S_n^\pm] = \pm S_{m+n}^\pm \qquad , \qquad [S_m^+, S_n^-] = -2S_{m+n+1}^0 \qquad (4)$$

Then, affine Lie algebra $SU(1,1)$ without central extension has been constructed by using these generators. The; Casimir operator is $\hat{C}_2 = S^0(S^0 - 1) - S^\pm$, too.

The IBM-1 Hamiltonian, without distinguishing between proton and neutrons, in this formalism, has written as:

$$\hat{H}_{regular} \equiv \hat{H}_{SU(1,1)} = gS_0^+ S_0^- + \varepsilon S_1^0 + \hat{C}_2[(\gamma SO(5) + \delta SO(3)] \ . \ (\varepsilon, \gamma, \delta \text{ and } g \text{ are real parameters}) \qquad (5)$$

$C_s = C_d$ corresponds with the SO(6) limit. Also, the U(5) limit has been described by the $c_s = 0$ requirements, and finally, $c_s \neq c_d$ has been used for the explanation of the U(5)↔O(6) transitional region. We present the regular or unperturbed states of $SU(1,1)$ algebra by with $x_i$ number parameters $c-$n terms of unknown, dwhich define $|N, k; v_s v n_\Delta LM\rangle$ as:, $i = 1, 2, ..., k$

$$|N, k; v_s v n_\Delta LM\rangle = \sum_{n_i \in Z} a_{n_1} a_{n_2} ... a_{n_k} x_1^{n_1} x_2^{n_2} ... x_k^{n_k} S_{n_1}^+ S_{n_2}^+ ... S_{n_k}^+ |lw\rangle \ , \qquad (6)$$

$k$ relate to the total number of boson $N$ as $N = 2k + v + v_s$. The analytical behavior of wave functions allows us to consider $x_i$ near zero. Now, the wave functions of $SU(1,1)$ algebra are yielded by using the commutation relations between the generators as:

$$|N, k; v_s v n_\Delta LM\rangle = N S_{x_1}^+ S_{x_2}^+ ... S_{x_k}^+ |lw\rangle \ , \qquad (7)$$

Where N is the normalization factor and

$$S_{x_i}^+ = \frac{c_s}{1 - c_s^2 x_i} S^+(s) + \frac{c_d}{1 - c_d^2 x_i} S^+(d) \ , \qquad (8)$$

Also, we defined the $x_i$ values via the following set of equations:

$$\frac{\varepsilon}{x_i} = \frac{gc_s^2(v_s + \frac{1}{2})}{1 - c_s^2 x_i} + \frac{gc_d^2(v + \frac{5}{2})}{1 - c_d^2 x_i} - \sum_{i \neq j} \frac{2}{x_i - x_j} \ , \qquad for \ i = 1, 2, ..., k \qquad (9)$$

The eigenvalues of Eq.5, regular Hamiltonian by 0p-0h excitation, are determined as:

$$E^{(k)} = \langle N, k; v_s v n_\Delta LM | \hat{H}_{SU(1,1)} | N, k; v_s v n_\Delta LM \rangle \ , \qquad (10)$$

The final form is:

$$E^{(k)} = h^{(k)} + \gamma v(v+3) + \delta L(L+1) + \alpha \Lambda_1^0$$
$$\Lambda_1^0 = \frac{1}{2}\left[c_s^2\left(v_s + \frac{1}{2}\right) + c_d^2\left(v + \frac{5}{2}\right)\right] \quad \text{and} \quad h^{(k)} = \sum_{i=1}^{k} \frac{\alpha}{x_i} \tag{11}$$

Now, we must extract the parameters of this equation in comparison with experimental data for energy levels of $^{112}$Cd nucleus. To this aim firstly, we selected, $\varepsilon = \frac{\varepsilon}{g}$, $g = 1$ keV, $c = \frac{c_s}{c_d} \leq 1$ similar to alautc values of ase wit rofhc $i=1$) have solved for8then, Eq. ( $y_i = c_d^2 x_i$ and what have done in Refs.[20]. After getting the roots for each level, and m identical to what that have done in Refs.[20]. $v$ and $v_s$ combination into $h^{(k)}$, we can use some extraction process such as least square fitting are other global methods but to get the exact results we used the roots of equations by beth-anthats methods, we follow the method that, used the least square fit in Matlab software to extract $\delta$ and $\gamma$ [21]. These processes, have been repeated by using the different values of $c$ and $\varepsilon$ to reduce the difference between theoretical predictions for different regular and intruder energy levels and experimental counterparts. We changed the $c_s$ values between 0 and 1 limits with $\Delta c_s = 0.1$ step length; determined all roots, parameters, and energy values for each case; some of these results have presented in Table1.We also used the

$$\sigma = \sqrt{\frac{1}{N^2}\sum_i (E_{\exp} - E_{Th})^2}$$, $N$, the number of considered levels, as the quality measure for our extracting procedure, which their values for different $c_s$ are present in Table 1, too.

Table1. Regular; and intruder energy levels of $^{112}$Cd nucleus have been determined by using different $c_s$ values with related quantum numbers in the framework of $SU(1,1)$ algebra. The experimental data are taken from Ref. [28]. $\sigma$ describes the quality of extraction processes. All; of the energy values are in keV.

| Level | k | v | $v_s$ | $E_{exp}$ | $E_{th}(c_s = 0)$ | $E_{th}(c_s = 0.2)$ | $E_{th}(c_s = 0.4)$ | $E_{th}(c_s = 0.6)$ | $E_{th}(c_s = 0.8)$ | $E_{th}(c_s = 1)$ |
|---|---|---|---|---|---|---|---|---|---|---|
| $0_1^+$ | 4 | 0 | 0 | 0 | 0 | 0 | 0 | 0 | 0 | 0 |
| $2_1^+$ | 3 | 2 | 0 | 617 | 594 | 580 | 600 | 741 | 801 | 875 |
| $0_2^+$ | 3 | 1 | 1 | 1224 | 1297 | 1311 | 1328 | 1362 | 1375 | 1411 |
| $2_2^+$ | 3 | 1 | 1 | 1312 | 1374 | 1785 | 1789 | 1801 | 1852 | 1940 |
| $4_1^+$ | 3 | 2 | 0 | 1415 | 1576 | 1800 | 1852 | 1889 | 1885 | 1912 |
| $0_3^+$ | 2 | 4 | 0 | 1433 | 1389 | 1274 | 1526 | 1621 | 1785 | 1796 |

| | | | | | | | | | |
|---|---|---|---|---|---|---|---|---|---|
| $2_3^+$ | 2 | 4 | 0 | 1468 | 1517 | 1540 | 1559 | 1567 | 1595 | 1507 |
| $0_4^+$ | 2 | 2 | 2 | 1870 | 1890 | 2088 | 2211 | 2321 | 2425 | 2611 |
| $4_2^+$ | 3 | 1 | 1 | 1871 | 1954 | 1980 | 1997 | 2007 | 2041 | 1962 |
| $3_1^+$ | 3 | 2 | 0 | 2065 | 2123 | 2310 | 2221 | 2450 | 2778 | 2913 |
| $4_3^+$ | 2 | 4 | 0 | 2082 | 2169 | 2105 | 2249 | 2271 | 2301 | 2381 |
| $2_4^+$ | 2 | 2 | 2 | 2122 | 2188 | 2209 | 2301 | 2285 | 2264 | 2350 |
| $2_5^+$ | 2 | 1 | 3 | 2156 | 2204 | 2270 | 2351 | 2384 | 2224 | 2370 |
| $6_1^+$ | 3 | 2 | 0 | 2168 | 2214 | 2454 | 2574 | 2679 | 2321 | 2376 |
| | | | $\sigma$ | 76.16 | 86.16 | 92.99 | 108.11 | 270.25 | 321.66 | |

Our results show good predictions of the transitional Hamiltonian, independent of $c_s$ values. We examine different values of $c_s$ and the best agreement between theoretical and experimental counterparts found in $c_s = 0.4$. This confirms our idea of using such transitional formalism as a regular part of this nucleus. These results are in agreement with the predictions of Jolie et al. [31-34] and Heyde et al. [31-32], which showed the advantages of U(5) dynamical symmetry in the description of the regular part in the energy spectra of the Cd isotopic chain, respectively shell model and a combination of U(5) and O(6) dynamical limits. Also, only for $4_2^+$ and $2_3^+$ as intruder levels, the distance between theoretical predictions and experimental values are apparent, and for other levels in ground and excited bands, this formalism makes appropriate results. Another; point in this table is the better results of the $c_s = 0$ case, which, as we have expected for the N bosons space, the U(5) limit excepted for such space and also described regular states with high accuracy. This result confirms the predictions of Lehman et al in Ref.[7,19], which used the Hamiltonian of U(5) limit for the regular states of this nucleus. In the following, we add new terms to our regular Hamiltonian and also suppose the N+2 bosons space in determining of the effects of these new terms to make an accurate description of intruder levels. This idea, adding the O(6) limit as a perturbation to the regular U(5) Hamiltonian, is similar to the method that Heyde et al. [31,32], have proposed because both groups have a standard O(5) subgroup.

2.2. Additional terms

The affine $SU(1,1)$ lie algebra, has been used here to provide a relevantly simple solvable pairing model that incorporates the mixing of two-particle and two-hole configurations. This transitional Hamiltonian in our calculations in the previous subsection works on the states of N bosons space described as $|N,k;v_s v n_\Delta LM\rangle$. Now, we introduce the method based on the extension of this transitional Hamiltonian by adding the Casimir operator of O(6) dynamical limit and also carrying out the calculation in the N+2 space, which the total wave function with the $|N+2,k;v_s v n_\Delta LM\rangle$ type. In this way, we followed the method developed by Lehman and Jolie, with the nature of $4_2^+$ and $2_3^+$ states, as have been predicted by Heyde et al. [22,31,32]. Rowe, in Refs. [22,24] has claimed that $\hat{S}^+\hat{S}^-$ the operator in the framework of $SU(1,1)$ algebra is the Casimir operator of O(6) symmetry. We extended our approach by adding the $\hat{S}^+\hat{S}^-$ term to affine $SU(1,1)$ algebra as 2p-2h excitation, similar to the method done by Lehman in Ref.[18]; In the framework of U(6) algebra. Another term that must add to our extended Hamiltonian is the mixing term which combines both N and N+2 bosons spaces. To this aim, we used the usual mixing Hamiltonian, which is defined in different papers as:

$$\hat{H}_{mix} = \eta[s^+ \times s^+ + s \times s]^0 + \chi[d^+ \times d^+ + d \times d]^0 \quad (12)$$

We also used that our considered transitional Hamiltonian connects both U(5) and O(6) dynamical limits by the variation of a control parameter of transitional Hamiltonian and, therefore; has the same role. The Hamiltonian to illustrate mixing could be written as follows:

$$\hat{H}_{tot} = P_N(\hat{H}_{SU(1,1)})P_N + P_{N+2}(g\hat{C}_1(O(6)) \equiv g\hat{S}^+\hat{S}^-)P_{N+2} + P_N(\hat{H}_{SU(1,1)} + \hat{H}_{mix})P_{N+2} \quad (13)$$

The; $P_N$ is the projection operator that represents the N-boson subspaces, while $P_{N+2}$ projects the subspaces to N + 2 bosons [18-19].

To determine the energy levels for these new extended Hamiltonian, we used again Eq.(10), where the considered $|\psi_i\rangle$ states to get eigenvalues are yielded by combining the two wave functions as:

$$|\psi_i\rangle = \kappa|\psi_N\rangle + \kappa'|\psi_{N+2}\rangle.$$

The eigenvalues of this considered perturbed Hamiltonian would as $\langle H_{tot}\rangle = E\delta_{N,N+2} + \langle\psi_N|H_{mix}|\psi_{N+2}\rangle$ whereAlso, we . is the unperturbed energy $E\delta_{N,N+2}$ determined the expectation values of the H$_{mix}$ through $\begin{pmatrix} H_{11}^{mix} & H_{12}^{mix} \\ H_{21}^{mix} & H_{22}^{mix} \end{pmatrix} \begin{pmatrix} \psi_N \\ \psi_{N+2} \end{pmatrix} = E_i \begin{pmatrix} \psi_N \\ \psi_{N+2} \end{pmatrix}$ equation and the results have presented in Table 2. To get the results in N+2 bosons space, the quantum numbers changed as $N+2=2k + v_s + v$, and therefore, the prediction of regular states was modified, too. In this stage, we have determined energy spectra of $^{112}$Cd nucleus via the expectation values of i) combination of regular Hamiltonian, $H_{SU(1,1)}$, together Casimir operator of O(6) dynamical symmetry and ii) total Hamiltonians, which introduced in Eq.(13). We do this, to show the effect of mixing term individually on the considered states and especially the intruder ones. Similar to what we have done for regular Hamiltonians, we have examined different values of $c_s$ in our calculation. The best agreement between theoretical predictions and experimental counterparts, is yield via $c_s = 0.5$ for both $SU(1,1)$ transitional Hamiltonians in regular, and mixed terms. Also, the $\kappa$ & $\kappa'$ values are listed in this table, too which describe the effect of N, and N+2 bosons spaces, respectively.

Table 2. Energy; spectra of the $^{112}$Cd nucleus were determined via the extended method. Parameters of total Hamiltonians in Eq.(13) are $g = 1.14$, $\alpha = 500$, $\gamma = 20.18$, $\delta = 32.99$, $\eta = 27.85$ and $\chi = 10.21$ (all in keV). Also; the best agreement is yield by $c_s = 0.5$ in both regular and mixed parts. Similar to Table 1, $\sigma$ describes the quality of extraction processes.

| Level | $k$ | $v$ | $v_s$ | $\kappa$ | $\kappa'$ | $E_{exp}$ | $\langle H_{regular}\rangle + \langle \hat{C}_1(O(6))\rangle$ | $\langle H_{tot}\rangle$ |
|---|---|---|---|---|---|---|---|---|
| $0_1^+$ | 5 | 0 | 0 | 0.99 | 0.14 | 0 | 0 | 0 |
| $2_1^+$ | 4 | 2 | 0 | 0.96 | 0.28 | 617 | 655 | 628 |
| $0_2^+$ | 3 | 2 | 0 | 0.91 | 0.41 | 1224 | 1285 | 1260 |
| $2_2^+$ | 3 | 4 | 0 | 0.81 | 0.59 | 1312 | 1345 | 1360 |
| $4_1^+$ | 4 | 2 | 0 | 0.87 | 0.49 | 1415 | 1506 | 1545 |
| $0_3^+$ | 3 | 3 | 1 | 0.79 | 0.61 | 1433 | 1438 | 1486 |
| $2_3^+$ | 3 | 4 | 0 | 0.76 | 0.65 | 1468 | 1482 | 1497 |
| $0_4^+$ | 3 | 2 | 2 | 0.64 | 0.77 | 1870 | 1758 | 1701 |
| $4_2^+$ | 3 | 4 | 0 | 0.37 | 0.93 | 1871 | 1893 | 1914 |
| $3_1^+$ | 4 | 2 | 0 | 0.81 | 0.59 | 2065 | 2012 | 2245 |
| $4_3^+$ | 3 | 3 | 1 | 0.61 | 0.79 | 2082 | 2147 | 2105 |
| $2_4^+$ | 3 | 2 | 2 | 0.57 | 0.82 | 2122 | 2163 | 2149 |
| $2_5^+$ | 3 | 1 | 3 | 0.52 | 0.85 | 2156 | 2198 | 2170 |
| $6_1^+$ | 4 | 2 | 0 | 0.74 | 0.67 | 2168 | 2200 | 2431 |
| | | | | | $\sigma$ | | 46.39 | 22.76 |

A detailed description of the selected energy levels, experimental values together theoretical predictions of regular, mixed, and total Hamiltonians are presented in Figure 1.

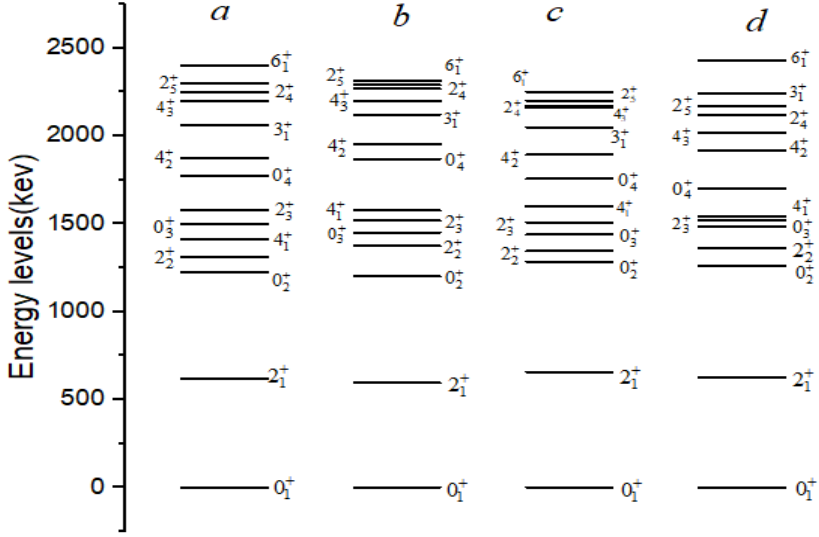

Figure 1. Energy; spectra of $^{112}$Cd nucleus. a); The experimental values, b) theoretical prediction $SU(1,1)$-based transitional Hamiltonian, which yield by $c_s = 0$, c) predictions of total Hamiltonian, and finally d) theoretical predictions of transitional Hamiltonian together O(6) Casimir operator which the two latest are yield by $c_s = 0.5$.

The extended Hamiltonian, Eq.(13), obviously improves theoretical predictions for energy spectra of $^{112}$Cd. These corrections are apparent for the regular states while for intruder levels, $4_2^+ \,\&\, 2_3^+$, this approach completely changes our previous results. The comprehensive process suggests more exact results for the levels in a ground band compared with the excited bands, too. Also, the accuracy of this extended formalism decreased for high spin levels, even for J = 6, and one may conclude that we must consider the sdg-IBM version of such comprehensive models. These results confirm the idea of symmetry mixing and extension of Hamiltonians in describing intruder structures. On the other hand, for intruder states, the results of column 8 of Table 2, which yield the expectation values of transitional Hamiltonian together O(6) Casimir operator, show better agreement. This may relate to the nature of these levels, which have been reported in Ref.[19]. And also, our results for $\kappa\,\&\,\kappa'$ coefficients verify it. The weight of N and N+2 bosons space, described by these coefficients, give exciting results about regular and intruder levels, too. For regular states in the ground band, the N boson space, which corresponds with the U(5) dynamical limit, has a prominent role, and the impact of this dynamical limit has reduced when the spin of the considered state increases. For other levels in the excited bands, the effect of N+2 boson space increased, and the maximum values of $\kappa'$ yield for $4_2^+ \,\&\, 2_3^+$ intruder states. In the next section, we consider the quadrupole transition rates and examine the effect of considering wave functions as combining two different boson spaces on theoretical predictions for these quantities.

## 3. E2 Transition probabilities

Electromagnetic transition rates are known as the most sensitive observable to a nuclear structure, and their analyses make remarkable data on the mixing of symmetries [19,22,31,32], shape coexistence [16,19], and partial dynamical symmetry [19,33].The; quadrupole transition rates are determined by using the following equation:

$$B(E2, L_i \to L_f) = \frac{1}{2L_i + 1} |\langle \psi_f | T(E2) | \psi_i \rangle|^2 \tag{14}$$

The most general operator within the IBM-1 framework for E2 transition is [25]:

$$T(E2) = e\left[(s^+ \times d + d^+ \times s)^2 + \chi(d^+ \times d)^2\right], \tag{15}$$

The selection rules for two terms of this operator are $\Delta n_d = \pm 1$ and $\Delta n_d = 0$, respectively. In this section, we calculated such transition rates of $^{112}$Cd nucleus whose experimental data are available [19] and happened between our considered states, e.g., up to $6_1^+$ level. We considered two ways in our calculation, and compared the effect of different bosons spaces on the theoretical results. In the first one, all of the regular and intruder states have been labeled in the only N boson space, and therefore, we considered wave functions of Eq.(14) as statestion of these he second approach, the wavefun n tI . $|\psi_i\rangle = |\psi_N\rangle \equiv |N, k; v_s v n_\Delta L M\rangle$ described as the combination of N and N+2 bosons spaces, e.g. $|\psi_i\rangle = \kappa |\psi_N\rangle + \kappa' |\psi_{N+2}\rangle$. Also, we used the $\kappa$ & $\kappa'$ values listed in Table 2. The effective charge, $e$, and the dimensionless quantity, $\chi$, in Eq.(15) are extracted in comparison with experimental data, too, whose values are $e$ = 1.857 W.u. and $\chi$ = - 0.17. The; prediction of these two approaches have presented in Table 3.

Table 3. The; experimental quadrupole transition probabilities with theoretical predictions. B(E2)$_N$;is the theoretical predictions that yielded by using the considered states in N boson space, and B(E2)$_{mix}$ is determined by using the wave functions as the combination of two N and N+2 boson spaces.

| Transition | B(E2)$_{exp}$ | B(E2)$_N$ | B(E2)$_{mix}$ |
|---|---|---|---|
| $2_1^+ \to 0_1^+$ | 30.3 | 36.2 | 32.5 |
| $0_2^+ \to 2_1^+$ | 51.0 | 49.1 | 49.6 |

| Transition | | | |
|---|---|---|---|
| $2_2^+ \rightarrow 2_1^+$ | 39.0 | 42.7 | 41.3 |
| $2_2^+ \rightarrow 0_1^+$ | 0.65 | 0.73 | 0.69 |
| $4_1^+ \rightarrow 2_1^+$ | 63.0 | 71.9 | 68.8 |
| $0_3^+ \rightarrow 2_2^+$ | 99.0 | 113.1 | 104.8 |
| $0_3^+ \rightarrow 2_1^+$ | 0.012 | 0.017 | 0.014 |
| $2_3^+ \rightarrow 0_2^+$ | 120.0 | 138.1 | 127.6 |
| $2_3^+ \rightarrow 0_1^+$ | 0.88 | 1.09 | 0.97 |
| $3_1^+ \rightarrow 4_1^+$ | 25.0 | 30.4 | 27.2 |
| $3_1^+ \rightarrow 2_2^+$ | 1.80 | 2.09 | 1.97 |
| $4_3^+ \rightarrow 2_3^+$ | 59.0 | 68.4 | 62.7 |
| $4_3^+ \rightarrow 0_3^+$ | 24.0 | 29.6 | 26.5 |
| $4_3^+ \rightarrow 2_2^+$ | 58.0 | 76.6 | 66.2 |
| $4_3^+ \rightarrow 2_1^+$ | 0.90 | 1.13 | 0.97 |
| $2_4^+ \rightarrow 0_3^+$ | 25.0 | 31.1 | 29.4 |
| $2_4^+ \rightarrow 2_2^+$ | 5.30 | 6.27 | 5.88 |
| $2_4^+ \rightarrow 2_1^+$ | 2.20 | 2.58 | 2.41 |
| $2_4^+ \rightarrow 0_1^+$ | 0.017 | 0.023 | 0.020 |
| $2_5^+ \rightarrow 2_3^+$ | 23.0 | 29.8 | 27.2 |
| $2_5^+ \rightarrow 2_1^+$ | 0.06 | 0.084 | 0.072 |

Theoretical results in both approaches suggest good agreement with their experimental counterparts and this verifies our selection for quantum numbers and also shows the accuracy of extraction processes. It was evident that when we consider the wave functions as the combination of states in both N and N+2 bosons spaces, the deviation between theoretical results and experimental values reduce obviously. The accuracy of these predictions is more for contraband transitions in comparison with inter band transition in both approaches. Also; the results of mixed formalism are so apparent for such transitions, which originated from or ended in intruder states.

The latest subject which we considered in this investigation is the test of a method that consists of extending wave functions of intruder states to describe transition rates with high exactness. To; this aim, we used two $2_3^+ \rightarrow 0_1^+$ and $2_3^+ \rightarrow 0_2^+$ transitions and considered the state as:

$$|2_3^+\rangle = \lambda|0_1^+\rangle + \lambda'|0_2^+\rangle$$

By using this definition in Eq.(14) and also $\lambda = 0.76$ and $\lambda' = 0.65$, which are yielded via extraction in comparison to experimental values, the probability of these transitions are yield as $B(E2, 2_3^+ \to 0_2^+) = 127.6$ and $B(E2, 2_3^+ \to 0_1^+) = 0.99$. One; may conclude the equivalency these two different methods in the calculation of such quadrupole transition rates due to the results yield for $\lambda$ & $\lambda'$ which are similar to the values of $\kappa$ & $\kappa'$ for this intruder state and also the exact predictions for transition rates. These; results with better predictions of mixed Hamiltonian for the energy spectra in comparison with the $SU(1,1)$ transitional Hamiltonian show the necessity to use mixed formalisms for the accurate description of regular and intruder states and also, transition rates.

## 4. Conclusions;

In this study, we introduced an extension of $SU(1,1)$- based Hamiltonian, for a detailed description of energy levels and quadrupole transition rates of the $^{112}$Cd nucleus. This considered transitional Hamiltonian suggests good predictions only for regular states, but for the intruders, ones need some corrections. The ability of considered Hamiltonian to coverage both U(5) and O(6) dynamical limits with standard O(5) sub algebras of them make it possible to extend this formalism by adding the Casimir operator of O(6) algebra and also mixing term. We considered the wave functions of both regular and intruder states as a combination in the N and N+2 boson spaces. The results of mixed formalism show improvement in the predictions for energy spectra and quadrupole transition rates. We would test the same procedure in the following research by adding the Casimir operator of U(5) dynamical limit for such nuclei located near the Z = 82 closed shell, and their intruder states have spherical nature.

**Acknowledgment;**

This work is supported by the Research Council of the University of Tabriz.

**Author contributions;**

M. Rastegar, H. Sabri, and A. O. Ezzati performed the initial calculations, analyzed and interpreted the results, and wrote the main manuscript text. All authors commented on and reviewed the manuscript.

**Competing interests;**

The authors declare no competing interests.